\documentclass[12pt,twoside]{article}
\usepackage{fleqn,espcrc1,epsfig}

\def\rsim{\mathrel{\raise2pt\hbox to 8pt{\raise -5pt\hbox{$\sim$}\hss{$>$}}}}
\def\lsim{\mathrel{\raise2pt\hbox to 8pt{\raise -5pt\hbox{$\sim$}\hss{$<$}}}}
\newcommand{\boldtau}{\mbox{\boldmath $\tau$}}

\begin {document}

\title{Three-Nucleon Forces for the New Millennium}

\author{J.\ L.\ Friar\address{Theoretical Division, Los Alamos National 
Laboratory, Los Alamos, New Mexico, 87545}}

\maketitle

\begin{abstract}
Most nuclear physics ranges from insensitive to relatively insensitive to
many-nucleon forces. The dominant ingredient in calculations of nuclear
properties is the nucleon-nucleon potential. Three-nucleon forces nevertheless
play an important role in nuclear physics because of the great precision of
modern calculational methods for systems of relatively few nucleons. We explore
the reasons why many-body forces are weak in nuclei by using a classification
scheme for such forces that is based on dimensional power counting, which is
used to organize chiral perturbation theory. An assessment will be made of how
close we are to a ``standard'' three-nucleon force. Recent advances in
determining the significance of three-nucleon forces will also be discussed.
\end{abstract}

\section{INTRODUCTION}

The turn of the century is a good time to assess the importance and impact of
three-nucleon forces (3NFs) on the development of the field of few-nucleon
physics.  It has been 67 years since Wigner\cite{wigner} first raised the
possibility that three-nucleon forces might be significant in the triton: ``
$\ldots$ one must assume a certain potential energy $\ldots$ or a three-body
force.''  It is significant that the triton had not yet been discovered,
although he predicted it would be bound by nucleon-nucleon (NN) forces alone. 
Since that time we have relied on field-theoretic techniques, phenomenology, and
sophisticated symmetry arguments to construct 3NFs, and the most modern and
advanced experimental facilities have recently been used to validate these
forces.

In a very real sense we are fortunate that three-nucleon potentials are not too
strong or too weak.  Indeed, I wouldn't be giving this talk if they were.
Imagine these forces to be $1-2$ orders of magnitude stronger than they actually
are. In that case 3NFs would be comparable to NN forces and (without stretching
the imagination) 4NFs, $\ldots$ could also be comparable.  In this scenario
nuclear physics would be intractable, and in all likelihood this conference
would not be held.  On the other hand we could imagine such forces to be $1-2$
orders of magnitude weaker than they actually are.  In this case they would play
almost no role in nuclear physics, and although few-nucleon physics would be
healthy and this conference would be held, the topic would be vacuous and there
would be no such talk.

\section{THREE-NUCLEON FORCE SCALES}

What sets the scales such that our universe lies between these limits, where
3NFs are weak but significant?  The answer lies in the scales associated with
QCD, which I will introduce later.  For better or worse, these scales allow me
to stand before you today and discuss these most interesting of forces.  Indeed,
these scales allow a qualitative discussion of many aspects of few-nucleon
physics, and I will rely on this approach to find common ground.

My first task is to estimate the size of the effect of three-nucleon forces 
using scales.  This can be achieved by a handwaving argument that is 
nevertheless correct in its essence.

\begin{figure}[htb]
\epsfig{file=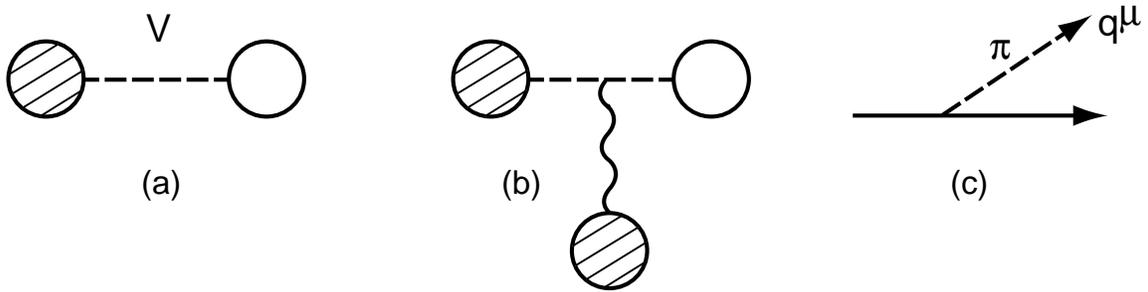,height=1.5in,clip=}
\caption{Generic nuclear force (dashed line) is shown in (a), while the
additional effect of a third nucleon is indicated in (b) by a wiggly line, and
the emission by a nucleon of a pion with four-momentum $q^{\mu}$ is illustrated 
in (c).}
\end{figure}

Figure~(1a) shows two nucleons interacting via an NN potential, $V$ (dashed
line). Adding another nucleon makes this a three-body system (Fig.~(1b)) and,
in addition to the normal NN interaction between the original two nucleons, that
force $V$ will somehow feel the effect of the additional nucleon (wavy line),
and the size of this additional effect on the energy should scale as $ V \cdot V
= V^2$, since all of the nucleons are the same.  This quantity unfortunately no
longer has the dimensions of energy and we need to divide by an additional
energy scale in order to obtain a final estimate.  We motivate this scale in
Fig.~(1c) by showing a virtual pion (with four-momentum $q^\mu$) emitted by a
nucleon. Normally we ignore the time component of $q^\mu (q^0)$, which scales
as the difference of kinetic energies of the initial and final nucleon; that
is, it scales as $1/M$, where $M$ is the nucleon mass.  Thus we might suspect
that the {\bf additional} effect of the third nucleon on the potential energy of
the original pair of nucleons scales as
\begin{equation}
\Delta V \sim \frac{V^2}{M c^2} \ \ ,
\end{equation}
where we have placed a factor of $c^2$ to make the dimensions correct. This
simple result, which in chiral perturbation theory has $M c^2$ replaced by
$\Lambda \sim 1$ GeV (a generic large-mass QCD scale) gives us a quick estimate
of the energy shift. Using $\langle V \rangle \sim 20-30$ MeV/pair we find
$\langle \Delta V \rangle \sim \frac{1}{2} - 1$ MeV, for what is either a
three-nucleon force effect, a relativistic effect (because of the $1/c^2$), or
an off-shell effect (this latter is not obvious, but is intimately related to 
the $q^0$ in our ``derivation''; it is an essential part of the
``quasipotential'' \cite{offshell} problem).

One important and obvious caveat for theorists is that it will be difficult to
interpret calculations that have numerical errors greater than $1$ MeV, if our
goal is to understand three-nucleon forces.  Indeed, we should do much better
than that and restrict triton errors to $ \sim 0.1 $ MeV $\sim 1\%$ of the
triton binding energy.  In addition, $1 \%$ absolute experiments are extremely
difficult and rare.  Calculations with numerical errors $\lsim 1\%$ have
consequently become the standard and are called ``exact,'' ``complete,'' or
``rigorous''.  They are one of the biggest success stories in our field in the
past 50 years.
\baselineskip 14pt

\section{THREE-NUCLEON SYSTEMS AND CALCULATIONS}

A bit of history is always a good way to start a discussion about the future. As
scientists we naturally tend to concentrate on our unsolved problems, and
successes are often overlooked.  In the process of giving my views on where the
field is going, I will also enumerate a few of the many successes in our
business, which highlight the progress that we have made\cite{walter,joe}.

One can conveniently categorize few-nucleon calculations as follows:  (a) bound
states (i.e., $^3$H and $^3$He); (b) Nd scattering below deuteron-breakup
threshold; (c) Nd scattering above deuteron-breakup threshold; (d) transitions
between bound and continuum states.  All of these types of calculations have
been performed, and benchmarked comparisons between different methods exist for
all categories except pd scattering (i.e., including a Coulomb force between
protons\cite{Levin}) at finite energies.  The ability to perform these extremely
difficult calculations, especially the scattering calculations, has been one of
the major successes in few-nucleon physics.  When one considers this together
with the incredible accomplishments of Vijay Pandharipande\cite{Vijay} and his
collaborators (including my colleague, Joe Carlson) for $A > 3$, this area is
one of the most successful in all of nuclear physics, and goes far beyond even
the dreams that theorists had 25 years ago.

I summarize this part of the talk by noting that $1 \%$ calculations are needed
in order to disentangle systematically the relatively small effects of
three-nucleon forces (or relativistic effects, off-shell effects, $\ldots$).
Such calculations are now possible using many different techniques.  Most
observables agree very well with experiment; indeed, most are insensitive to
3NFs.  The trick is to find the proper observables to investigate.

\section{HISTORY OF THREE-NUCLEON FORCES}

Wigner's mention of three-nucleon forces in his calculation of $^3$H was
subsequently ignored. There was no hope then (and little now) of being able to
calculate and interpret results for a strongly interacting system dominated by
such forces.  While we have significant and extensive experimental information
on the NN force, we have very little knowledge with which to constrain
three-nucleon forces.  We are forced to rely almost entirely on theory,
particularly on field theory. Early efforts involved primitive models that have
not left their mark on the field.  One calculation that has left an indelible
mark is the august Fujita-Miyazawa model\cite{FM}, based entirely on isobar
intermediate states and pion propagation within nuclei.  This was motivated by
the dominant role of nuclear resonances in some processes.

\begin{figure}[htb]
\epsfig{file=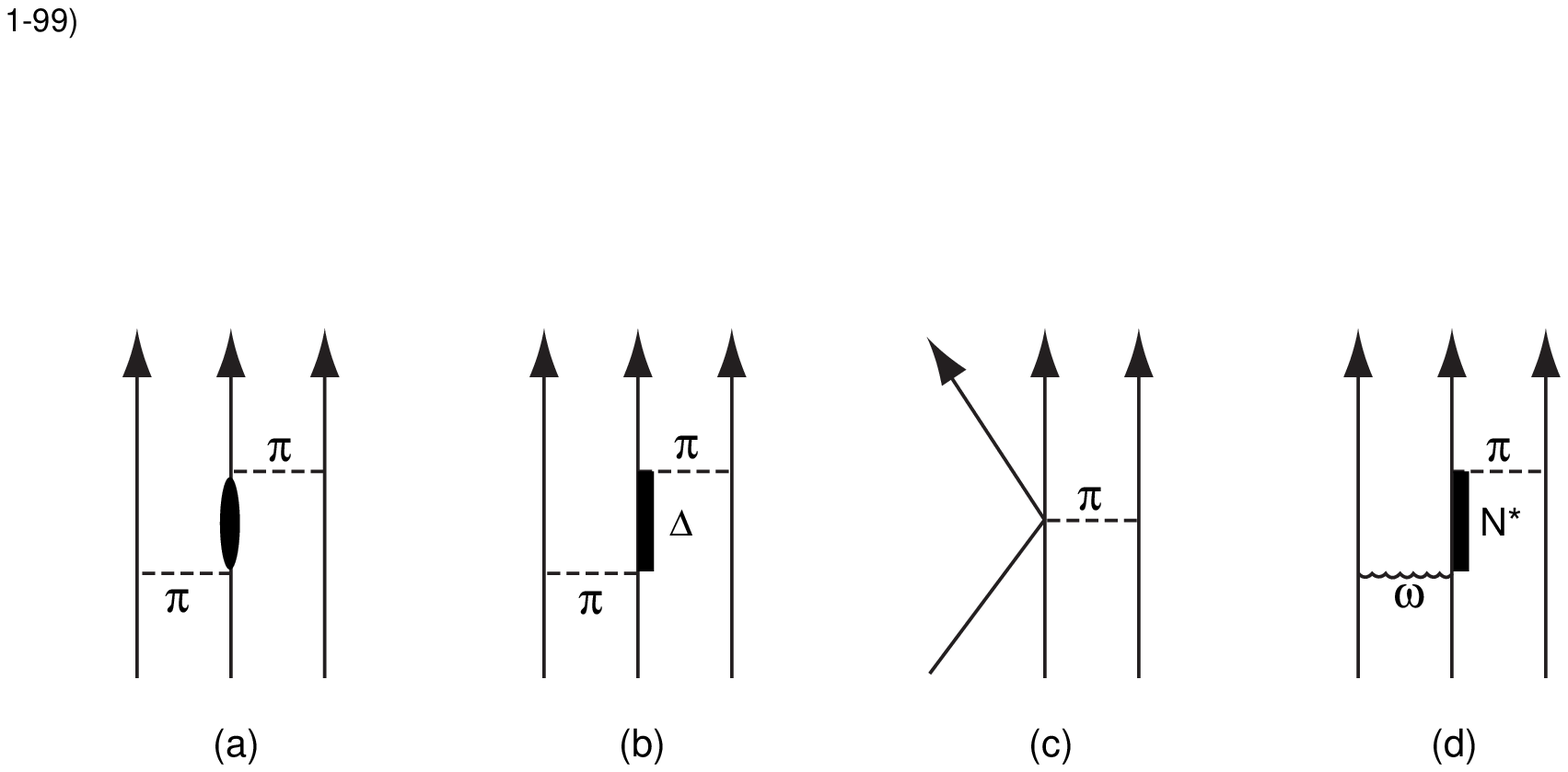,height=1.75in,clip=}
\caption{Mechanisms that contribute to three-nucleon forces. Two-pion-exchange
forces are shown generically in (a), and the important isobar contribution in 
(b). Chiral perturbation theory predicts a large contribution of the type shown 
in (c), a specific mechanism of that type being displayed in (d).}
\end{figure}

Examples both familiar and unfamiliar are shown in Fig.~(2).  In Fig.~(2a) a 
pion emitted by one nucleon interacts in a complicated way with a second nucleon
and then is absorbed by a third nucleon.  The Fujita-Miyazawa (FM) approximation
to the entire process is shown in Fig.~(2b).  One can also replace one (or both)
pions by a heavy meson (as shown in Fig.~(2d)).  A variety of such short-range
mechanisms are depicted in chiral perturbation theory by Fig.~(2c), where all of
the short-range processes (induced by heavy-mass particles) are shrunk to a
point.  We will first discuss the $2\pi$-exchange forces.

The second noteworthy calculation was performed by one of our conference
organizers, Shin Nan Yang\cite{Yang}.  The Yang model was the first
three-nucleon potential model based on chiral-symmetry considerations, although
there were previous calculations of effects based on PCAC. This model was used
in a variational calculation to estimate $\langle \Delta V_{\rm 3NF} \rangle
\sim 2$ MeV.  This set the stage for the development of the most widely used
3NF, the Tucson-Melbourne (TM) model\cite{TM}, which exploited current algebra
and PCAC (and treated off-shell effects in a serious manner) in deriving that
force. The various parameters of this model incorporate phenomenology (including
isobars) in a more meaningful way than the FM model. In addition to the models 
already mentioned, there are the Urbana-Argonne (UA)\cite{UA} [an offshoot of 
FM], the RuhrPot\cite{RuhrPot}, the Brazil\cite{Brazil}, and the 
[$\chi$PT-based] Texas models\cite{texas}.

It is not a good thing to have so many different models for what should be a
single correct physical force.  Indeed, one of our tasks for the new millennium
is to force a ``convergent evolution'' on these models by incorporating proper
amounts of the correct physics.  Recent progress has been made in that
direction.  The UA model now incorporates a chiral-symmetry-breaking term (the
``a'' term of TM) while a direct appeal to $\chi$PT\cite{3NF} leads to the
elimination of a rather unimportant term of short-range + pion-range character
(the ``c'' term) in the TM model, leaving the $2\pi$-exchange parts of the two
models essentially equivalent except for minor differences in parameter values. 
Thus $\chi$PT has produced a recent unification of $2\pi$-exchange forces.

We emphasize that in order to accomplish this we must incorporate two key pieces
of physics:
\begin{itemize}
\item adequate phenomenology (such as isobars)

\item chiral constraints .
\end{itemize}

\section{QCD AND ${\chi}$PT}

Before discussing the road to the future for 3NFs that have a short-range
component, it is necessary to implement an organizational scheme.  There are
simply too many possible operators that can contribute.  Chiral perturbation
theory fortunately allows us to sort 3NFs into classes and identify the terms
that should be dominant.  Indeed, this sorting process is the organizational
scheme of $\chi$PT.  How does it work?

The ``natural'' degrees of freedom of QCD are quarks and gluons, whose
interactions manifestly reflect the symmetries of QCD.  We are not required to
use these degrees of freedom, however, and the traditional degrees of freedom of
nuclear physics, namely nucleons and pions, are the most effective ones.  If we
imagine somehow mapping QCD expressed in terms of quarks and gluons onto the
Hilbert space of all particles, and then freezing out the effect of the heavy
particles (e.g., all nucleon resonances and all mesons with mass $\Lambda \rsim
1$ GeV) we arrive at chiral perturbation theory\cite{Weinberg}.  The
freezing-out process is familiar in nuclear physics as Feshbach [P,Q] reaction
theory\cite{Feshbach}, and is known in that context to lead to complicated
operators. It is nevertheless possible to implement the important
chiral-symmetry constraints of QCD in this ``QCD in disguise'' theory. Even more
important is the power counting that makes this scheme work as a field
theory\cite{Georgi}.

Power counting is a kind of dimensional analysis based on (only) two
energy scales associated with QCD.  One scale is $f_\pi$, the pion (beta-)decay
constant ($\sim 93$ MeV) that controls the Goldstone bosons (such as the pion),
while the second is $\Lambda \sim 1$ GeV, the scale of QCD bound states (the
$\rho$ and $\omega$ mesons, nucleon resonances, etc.), above which we agree to
freeze out all excitations.  That these scales are all that is needed is not
only not obvious, but it's a little bit miraculous!  Using these scales it can
be shown that a given (Lagrangian) building block should scale as
\begin{equation}
{\mathcal {L}}^{(\Delta)} \sim \frac{c}{f_\pi^\beta \Lambda^\Delta} \ 
{\rm (times
\  various \ fields)} \ \ .
\end{equation}
Two vital properties of this simple construction are:  (1) $\Delta \geq 0$
because of chiral symmetry; (2) c is a dimensionless constant that satisfies
$|c| \sim 1$, which is the condition of ``naturalness''. The latter is also not
obvious, but is extremely important.  If $|c|$ could vary over many orders of
magnitude in a problem, this organizational scheme would be useless.  Moreover,
unless positive powers of $\Lambda$ exist in the denominator (even in the
presence of vacuum fluctuations) this would not lead to an expansion in powers 
of (small/large).

This formal scheme can be implemented in nuclei to estimate the sizes of various
types of operators in the nuclear medium\cite{power}. An additional nuclear
scale is needed in order to characterize the medium, and this is given by the
{\bf effective} nuclear momentum (or inverse correlation length):  $Q \sim m_\pi
c$, where $m_\pi$ is the pion mass.  Then it is possible to show that the 
one-pion-exchange NN potential satisfies\cite{power}
\begin{equation}
\langle V_\pi \rangle \sim \frac{Q^3}{f_\pi \Lambda} \sim 30 \ 
{\rm MeV/pair}\, ,
\end{equation}
and
\begin{equation}
\langle V_{\rm 3NF} \rangle \sim \frac{Q^6}{f_\pi^2 \Lambda^3} \sim 
\frac{\langle V_\pi \rangle^2}{\Lambda} \sim 1 \ {\rm MeV/triplet}\, ,
\end{equation}
and we have reproduced our previous result with $\Lambda \sim M c^2$. Note that
this size estimate applies only to the leading-order terms; smaller terms exist
that might be significant in special situations.

\section{STATUS OF 3NF CALCULATIONS}

There are 7 basic types of 3NFs in leading order.  Four are of two-pion range,
two are of mixed pion-range + short-range, and there is a class of short-range +
short-range forces.  Figure~2 shows several examples. The generic
two-pion-exchange force is given by Fig.~(2a) and can be broken down into the
``a'', ``b'', and ``d'' terms of the TM force\cite{TM}, plus the so-called Born
terms. The latter have been derived\cite{CF} but have never been used in their
entirety (there are many terms) in any $^3$H calculation. The two mixed terms
are those represented generically in Fig.~(2c), one specific mechanism in this
category being that of Fig.~(2d) (the so-called $d_1$-term).  These terms have
been investigated only once or twice\cite{dirk}. Finally, there are purely
short-range terms of the type incorporated in the UA 3NF.

In the near future it will be necessary to investigate thoroughly the importance
of this set. Most urgent are the Born terms. The local terms are almost
certainly unimportant compared to the isobar mechanism, but none of the nonlocal
terms have been incorporated into existing calculations.  A preliminary and not
wholly satisfactory set of calculations\cite{dirk} exists for the mixed
short-range + pion-range forces.  These need to be extended and refined.
Finally, it is by no means certain that the effects of these different forces
are entirely linear when added together (as indicated in Ref.\cite{dirk}). Thus
a lot of different calculations need to be performed in different combinations
and for as many different observables as possible.  Completion of this exercise
will indeed provide us with a solid base in this area from which we can extend
few-nucleon physics into the new millennium.

\section{EVIDENCE FOR THREE-NUCLEON FORCES}

We have postponed until the end a discussion of evidence for these forces, both
direct and indirect. The indirect evidence is strong but not compelling. With
all modern ``second-generation'' NN forces the triton is underbound by roughly 
$\frac{1}{2} - 1$ MeV, in agreement with our earlier estimate, and $^4$He,
$\ldots$ are also underbound. Unless our understanding of these NN forces is 
badly deficient, such NN forces require an additional three-nucleon force.

Better evidence is provided by a recent analysis of the tail of the $pp$
potential\cite{mart}, which generated very strong support for calculations of
two-pion-exchange forces obtained from $\chi$PT.  Several effective coupling
constants (for pion-nucleon scattering operators) were fit in that analysis, and
they agree with the same couplings that are used in two-pion-exchange 3NFs,
validating the latter mechanisms. In other words, once the building blocks have
been established it makes little difference what those blocks are used to
construct.

Direct evidence is available in the Sagara discrepancy, which is shown in 
Fig.~(3) in the differential cross section for pd scattering at 65
MeV\cite{sagara}. If we ignore the effects of the Coulomb interaction in the
forward direction, agreement between the experimental data and the calculation
with NN forces alone (dashed line) is very good, except in the diffraction
minimum where the data lie above the calculation.  If one adds a 3NF the solid
curve results, which nicely fills in the minimum, and agreement with the data is
quite good. This is rather strong evidence for 3NFs and it exists at other
energies. An estimate of the effect of the 3NF alone that is based on DWBA is
shown in the long-dashed curve. This general behavior is very familiar in
Glauber scattering at high energy, where a dominant single-scattering term falls
rapidly with increasing angle, until the double-scattering amplitude (which
decreases more slowly with angle) becomes dominant, and so on.

\begin{figure}[htb]
\epsfig{file=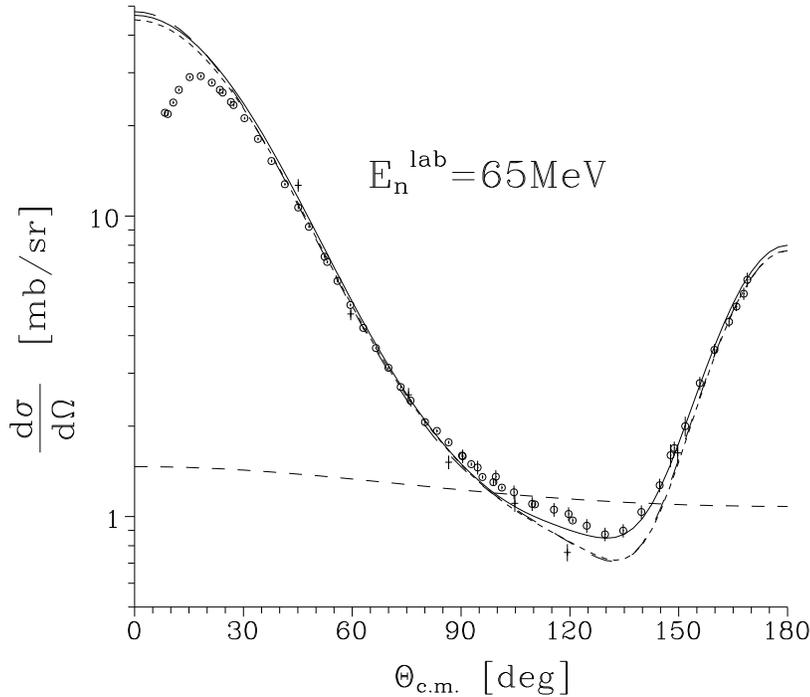,height=4.0in,bbllx=0pt,bblly=0pt,bburx=498pt,bbury=450pt}
\caption{Differential cross section for 65 MeV proton-deuteron scattering, 
showing calculations with $N$-$N$ forces only (dashed lines), a full calculation
that includes the TM 3NF (solid line), and an estimate of the effect of the 3NF 
alone (long-dashed line).}
\end{figure}

Finally, we discuss the effects of the short-range + pion-range terms, and in
particular the one depicted in Fig.~(2d). This is the so-called $d_1$-term that
has recently been shown\cite{dirk} to have a nonnegligible effect on the nucleon
asymmetry $(A_y)$ scattering observable.  That observable has been a serious
problem for theorists for a long time.  Calculations using NN forces alone are
significantly smaller than the data, particularly at low energies (e.g., $3$
MeV), both for the pd and nd scattering.  A variety of explanations have been
proposed (this observable is very sensitive to spin-orbit interactions), but the
most plausible is the 3NF mechanism\cite{A_y}.  An example of this is Nd
scattering at 3 MeV.  Calculations that incorporate only an NN force are about
30\% smaller than the experimental data at the maximum. Incorporating the TM 3NF
removes about $\frac{1}{4}$ of the discrepancy.  Adding the $d_1$-term (that we
discussed before) with a dimensionless strength coefficient $c_1 = -1$ removes
another $\frac{1}{4}$ of the discrepancy.  Technical problems prevented
calculations with stronger versions of this 3NF. It is nevertheless clear that
3NFs of various types make significant contributions to this observable. Much
more work is needed on this problem. We note that discrepancies also exists for
electromagnetic reactions and in the four-nucleon problem.

Special circumstances may dictate that classes of 3NF operators smaller than
leading order will play a role.  An example of this is neutron matter (or
neutron-rich nuclei). The isospin dependence of three-nucleon forces takes 3
forms:  $\boldtau_1 \cdot \boldtau_2 \times \boldtau_3$, $1$, and $\boldtau_i
\cdot \boldtau_j$. The first vanishes for three neutrons.  Because 3 neutrons
exist in a $T = \frac{3}{2}$ state, only the projection $(3 + \boldtau_i \cdot
\boldtau_j)/4$ contributes to that state, while the projection $(1 - \boldtau_i
\cdot \boldtau_j)/4$ vanishes. Some mechanisms (such as isobar configurations)
that prefer large isospins may be enhanced, as shown by Vijay Pandharipande and
his collaborators.  In these special circumstances the dimensionless isospin
factors (which typically average to about 1) can conspire to give enhanced
coupling strengths.  This has not yet been investigated in detail, but it should
be.

\section{SUMMARY AND PROGNOSIS}

We have made great advances in the past 25 years in our understanding of both
three-nucleon systems and three-nucleon forces.  We stand poised to make further
advances, based on recent technical developments.  Hopefully we will soon be 
able to develop a consensus ``standard model'' of 3NFs with all significant
features incorporated, which will then allow us to pursue three-nucleon physics
into the new millennium.

We summarize this section as follows.
\begin{itemize}
\item Most three-nucleon observables are insensitive to 3NFs.

\item 3NFs are small in size but appear necessary for the $^3$H binding energy,
the Sagara discrepancy, and the $A_y$ puzzle.

\item Chiral symmetry provides a unified approach to 3NFs; power counting
identifies dominant mechanisms.

\item The leading-order (dominant) 2$\pi$-exchange 3NFs have been calculated;
they have large isobar contributions.

\item New short-range plus pion-range mechanisms may help resolve the $A_y$ 
puzzle.

\item Although much remains to be investigated, a consensus appears to be
developing for the bulk of 3NF terms.

\end{itemize}

\section{ACKNOWLEDGEMENTS}

The work of J.L.F. was performed under the auspices of the United States 
Department of Energy.

\end{document}